\documentclass[letterpaper, journal]{IEEEtran}

\usepackage{cite}
\usepackage{graphicx}
\graphicspath{{fig/}}
\usepackage{amsmath}
\usepackage{algorithmic}
\usepackage{array}
\usepackage[caption=false,font=footnotesize]{subfig}
\usepackage{url}
\usepackage{lettrine}
\usepackage{amssymb}
\usepackage{amsthm}
\usepackage{multicol}
\usepackage{multirow}
\usepackage{mdwmath}
\usepackage{mdwtab}
\usepackage{makecell} 
\usepackage{tikz} 
\usetikzlibrary{shapes,arrows,calc}
\usepackage{float}

\begin{document}

\title{Connectivity Management in Satellite-Aided Vehicular Networks with Multi-Head Attention-Based State Estimation}
 
\author{Ibrahim Althamary, 
Chen-Fu Chou, 
and~Chih-Wei Huang,~\IEEEmembership{Member,~IEEE}
\thanks{This work is supported in part by the National Science and Technology Council of Taiwan under Grant 112-2221-E-008-059-MY2. (Corresponding author: Chih-Wei Hunag.)}
\thanks{Ibrahim Althamary is with the Interdisciplinary Research Center for Intelligent Secure Systems, KFUPM, Saudi Arabia (e-mail: ibrahim.thamary@kfupm.edu.sa).}
\thanks{Chen-Fu Chou is with the Department of Computer Science and Information Engineering, National Taiwan University, Taipei 106319, Taiwan (e-mail: ccf@csie.ntu.edu.tw).}
\thanks{Chih-Wei Huang is with the Department of Communication Engineering, National Central University, Taoyuan 320317, Taiwan (e-mail: cwhuang@ce.ncu.edu.tw).}
}

\maketitle

\begin{abstract}
Managing connectivity in integrated satellite-terrestrial vehicular networks is critical for 6G, yet is challenged by dynamic conditions and partial observability. This letter introduces the Multi-Agent Actor-Critic with Satellite-Aided Multi-head self-attention (MAAC-SAM), a novel multi-agent reinforcement learning framework that enables vehicles to autonomously manage connectivity across Vehicle-to-Satellite (V2S), Vehicle-to-Infrastructure (V2I), and Vehicle-to-Vehicle (V2V) links. Our key innovation is the integration of a multi-head attention mechanism, which allows for robust state estimation even with fluctuating and limited information sharing among vehicles. The framework further leverages self-imitation learning (SIL) and fingerprinting to improve learning efficiency and real-time decisions. Simulation results, based on realistic SUMO traffic models and 3GPP-compliant configurations, demonstrate that MAAC-SAM outperforms state-of-the-art terrestrial and satellite-assisted baselines by up to 14\% in transmission utility and maintains high estimation accuracy across varying vehicle densities and sharing levels.
 
\end{abstract}


\section{Introduction}

\lettrine{I}{ntegrated} satellite-terrestrial networks (ISTN), combined with multi-connectivity and sidelink technologies, are crucial for enhancing spectral efficiency and coverage in future 6G vehicular-to-everything (V2X) system\cite{He2021}. Efficient connectivity management is essential for optimizing resource allocation and ensuring seamless connectivity in dynamic vehicular environments\cite{shang2024multi,gupta2025pora}.

Traditional ISTN resource allocation methods have been extensively studied. Deng et al. \cite{deng2019joint} introduced multigroup precoding to enhance the capacity and quality of service (QoS) of ISTN; however, their method assumes low mobility and results in high computational complexity. Birabwa et al. \cite{birabwa2022service} applied genetic algorithms (GAs) for ISTN resource allocation; however, their approach relies heavily on comprehensive network information and iterative optimization, which limits its real-time applicability in dynamic V2X conditions. Hence, there is a critical need for more adaptive and efficient methods that can handle rapidly changing conditions while meeting V2X performance constraints.

Recent advances in multi-agent reinforcement learning (MARL) offer promising solutions for dynamic spectrum management in V2X contexts. Liang et al. \cite{liang2019spectrum} proposed deep Q-networks (DQN) to enhance spectrum sharing specifically in terrestrial vehicle-to-infrastructure (V2I) and vehicle-to-vehicle (V2V) links without considering vehicle-to-satellite (V2S) connectivity. Ji et al. \cite{ji2023multi} employed a dueling double DQN (D3QN-LS) with fingerprinting for terrestrial vehicular networks to enhance spectrum selection and power allocation. Chen et al. \cite{chen2023multi} introduced an actor-critic method with straightforward state estimation techniques. Nevertheless, integrating satellite connectivity remains challenging due to the need for distributed decision-making and robust state estimation in partially observable, dynamic vehicular scenarios.

Advances in machine learning, including multi-head attention (MHA) architectures \cite{vaswani2017attention}, fingerprinting \cite{foerster2017stabilising}, and self-imitation learning (SIL) with prioritized experience replay \cite{oh2018self,ji2023multi}, provide opportunities to improve state estimation and agent performance in complex, partially observable environments. These techniques significantly enhance both individual and collective decision-making capabilities in dynamic network scenarios.

This paper proposes the Multi-Agent Actor-Critic with Satellite-Aided Multi-head self-attention (MAAC-SAM) framework, designed explicitly for adaptive connectivity management across satellite, infrastructure, and vehicle-to-vehicle communication modes. It surpasses terrestrial-only systems, ensuring reliability in areas with sparse coverage. The framework uses MHA with gated recurrent unit (GRU) encoders for vehicles to share observations and achieve precise state estimation in partially observable environments, while self-imitation learning and fingerprinting improve learning efficiency and real-time decisions. The main contributions include the following.
\begin{itemize}
  \item To our knowledge, MAAC-SAM is the first to leverage MHA in MARL for satellite-aided vehicular networks, improving state estimation across heterogeneous V2X links. This integration robustly maintains accurate state estimation by effectively focusing on critical observation features in various observation sharing levels and vehicular densities.
  \item The MAAC-SAM framework uses SIL and fingerprinting to enhance learning and decision making in partially observable dynamic environments. SIL improves learning efficiency by reinforcing historically successful actions, while fingerprinting explicitly accounts for evolving policies of other agents, significantly improving adaptability.
   
  \item MAAC-SAM offers an adaptive strategy for vehicles to autonomously choose between V2V, V2I, or V2S modes and optimize resources such as subchannels and transmission power. This uses advanced state estimation and learning to ensure efficient spectrum use and robust network performance, significantly advancing autonomous decision-making in satellite-aided vehicular networks.
   
\end{itemize}
Simulations and ablation studies confirm that MAAC-SAM improves overall utilization compared to state-of-the-art baselines, demonstrating significant potential for satellite-aided V2X communication systems.

\section{System Model and Problem Formulation}\label{sec:system}
%


We consider a satellite-aided vehicular network consisting of vehicles \( \mathcal{I} = \{1,\dots,I\} \) that dynamically select among three connectivity modes: V2S, V2I, and V2V. Vehicles use \( K^V \) orthogonal S-band terrestrial sub-channels shared between V2I and V2V modes and \( K^S \) dedicated Ka-band sub-channels individually allocated for V2S communications. 

\begin{figure}
 \centering
 \includegraphics[width=\linewidth]{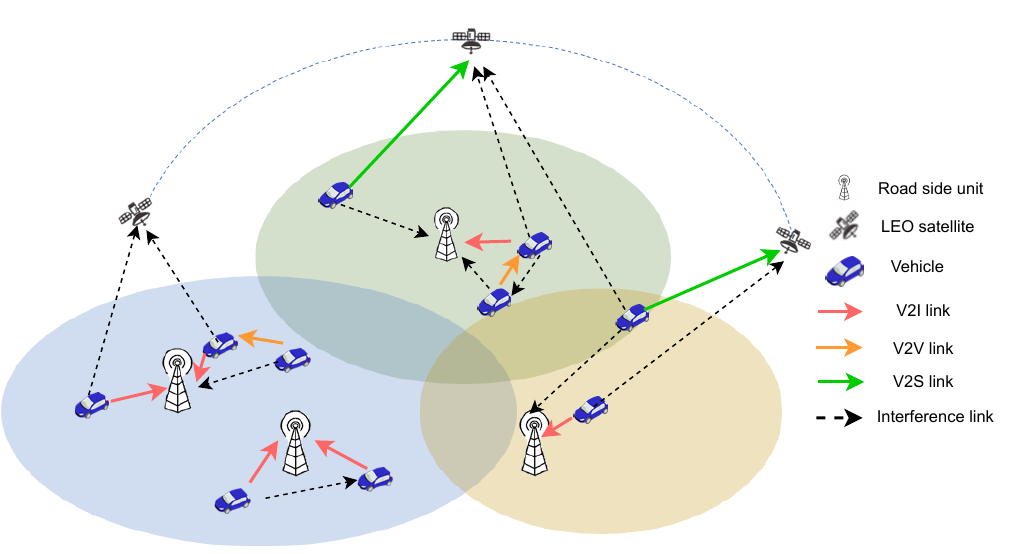}
 \caption{System model for the satellite-aided vehicular network.}
 \label{fig:model}
\end{figure}

As illustrated in Fig.~\ref{fig:model}, connectivity mode selection depends on vehicle position, channel conditions, and communication demands. Vehicles in the blue oval use terrestrial communications (V2V and V2I) due to their proximity to roadside units (RSUs). Those in the orange oval mainly rely on V2S and V2V when terrestrial coverage is sparse. Vehicles in the green oval can use all three modes (V2S, V2I, V2V) if positioning and channel conditions are optimal.

The total set of sub-channels is denoted as \( \mathcal{K}=\{1,\dots, K\} \), where \( K=K^V+K^S \), serving communication targets \( x \in \mathcal{X}=\{\text{RSU},\text{SAT},1,\dots, I\} \), representing RSUs, satellites (SAT), and vehicles, respectively.

The channel gain \( g_{x,k}^{i} \) from vehicle \( i \) to target \( x \) over sub-channel \( k \) explicitly incorporates the distance \( d_x^i \), path-loss exponent \( \kappa \), fading coefficient \( h_{x,k}^{i} \), and atmospheric attenuation \( A_a \), aligned with ITU-R standards~\cite{series2017propagation}:
 
\begin{equation}
g_{x,k}^{i} = (d_x^i)^{-\kappa}|h_{x,k}^{i}|^2,\quad
g_{\text{SAT},k}^{i} = \frac{G_T^i G_R^i}{A_a L_{\text{FS}}},
\end{equation}
where \( G_T^i \), \( G_R^i \), and \( L_{\text{FS}} \) denote the transmitter gain, receiver gain, and free-space loss, respectively.

The achievable transmission capacity \( C_t^{i} \) for vehicle \( i \) at time \( t \) is defined as:
\begin{equation}
C_t^{i} = B_{x_t^{i}}\log_2\left(1+\frac{P_t^{i}g_{x_t^{i},k_t^{i}}^{i}}{N_0+I_t^{i}}\right),
\end{equation}
where \( B_{x_t^{i}} \) is the allocated bandwidth, \( P_t^{i} \) is the transmission power, and \( N_0 \) is the noise power. Interference power \( I_t^{i} \) is explicitly calculated as:
\begin{equation}
I_t^{i} = \sum_{j\neq i}\mathbf{1}(k_t^j=k_t^i)P_t^{j}g_{x_t^{i},k_t^{i}}^{j}.
\end{equation}
For satellite communications (V2S), each vehicle is assigned a dedicated sub-channel, eliminating interference (\( I_t^i=0 \)), simplifying the capacity expression:
\begin{equation}
C_t^{i} = B_{\text{SAT}}\log_2\left(1+\frac{P_t^{i}g_{\text{SAT},k_t^{i}}^{i}}{N_0}\right).
\end{equation}

We define the utility function \( U^i \) to measure successful data transmission within latency constraints explicitly as:
\begin{equation}
U^i = \begin{cases}
1 & \text{if}\quad D_{\tau_L}^{i}\geq L^{i}, \\
0 & \text{otherwise},
\end{cases}
\end{equation}
where \( D_{\tau_L}^{i}=\sum_{t=1}^{T}C_t^{i}\cdot\Delta t \) is the cumulative data transmitted by latency deadline \( \tau_L=T \), and \( L^{i} \) is the required data size.

Our objective is to maximize overall transmission success across all vehicles:
\begin{equation}
\max_{\mathbf{X},\mathbf{K},\mathbf{P}}\sum_{i\in\mathcal{I}}U^i
\end{equation}
subject explicitly to practical system constraints:
\begin{equation}
0 \leq P_t^{i}\leq P_{\max},\quad\forall i\in\mathcal{I},\quad t=1,\dots,T,
\end{equation}
\begin{equation}
x_t^{i}\in\mathcal{X},\quad k_t^{i}\in\mathcal{K},\quad\forall i\in\mathcal{I},\quad t=1,\dots,T,
\end{equation}
with decision variables explicitly denoted as \(\mathbf{X}=\{x_t^i\}_{t,i}\), \(\mathbf{K}=\{k_t^i\}_{t,i}\), and \(\mathbf{P}=\{P_t^i\}_{t,i}\).
This formulation explicitly integrates realistic channel modeling, interference management, and practical resource constraints (power, bandwidth, latency). It enables adaptive connectivity management and efficient resource allocation in satellite-aided vehicular networks.

\section{Multi-Agent Reinforcement Learning for V2X}\label{sec:algorithm}

\subsection{POMDP Model for Networked Agents}

Vehicles in satellite-aided vehicular networks act as autonomous agents, dynamically coordinating connectivity based on partial observations of system states. Due to mobility, fluctuating channels, and decentralized decisions, the system is modeled as a partially observable Markov decision process (POMDP).

Formally, the POMDP is defined by the tuple \((\mathcal{G}, \mathcal{S}, \mathcal{O}, \mathcal{A}, \mathcal{T}, R)\). The communication graph \(\mathcal{G}=(\mathcal{I},\mathcal{E})\) includes vehicles as agents \(\mathcal{I}\) and feasible links \(\mathcal{E}\subseteq\mathcal{I}\times\mathcal{I}\) for V2V, V2I, and V2S connections. Global states \(\mathcal{S}\) reflect the system at each step, including positions, channels, and transmission progress. Joint observation space \(\mathcal{O}=(\mathcal{O}^1,\dots,\mathcal{O}^I)\) provides partial local insights per agent. Joint actions \(\mathcal{A}=(\mathcal{A}^1,\dots,\mathcal{A}^I)\) define possible agent decisions. The transition function \(\mathcal{T}:\mathcal{S}\times\mathcal{A}\to\mathcal{S}\) dictates probabilistic state evolution, and the reward \(R:\mathcal{S}\times\mathcal{A}\to\mathbb{R}\) assesses performance.

At each step \(t\), vehicle \(i\)'s observation \(o_t^i\) consists of previous-step SINR \(\gamma_{x,t-1}^i=(P_{t-1}^i g_{x^i_{t-1},k^i_{t-1}}^i)/(N_0+I_{t-1}^i)\) and remaining load \(\phi_{x,t}^i\) for modes \(x\in\mathcal{X}\):
\begin{equation}
o_t^i=\left(\{\gamma_{x,t-1}^i\}_{x\in\mathcal{X}},\{\phi_{x,t}^i\}_{x\in\mathcal{X}}\right).
\end{equation}
Vehicles dynamically exchange observations $\{o^j_t|j \in \mathcal{Z}_t(i)\}$ with neighboring agents $\mathcal{Z}_t(i)$ via sidelink control channels, where both neighbor sets and their sizes vary over time due to mobility.
Vehicle \(i\) selects action \(a_t^i=(x_t^i,k_t^i,P_t^i)\) based on \(o_t^i\), defining its resource allocation and communication strategy.
The reward function aligns explicitly with the utility in Section~\ref{sec:system}, incentivizing timely transmission completion:
\begin{equation}
r_t^i=U_t^i - w\cdot\frac{\phi_t^i}{L^i},
\end{equation}
where utility \(U_t^i\) indicates successful completion at time \(t\):
\begin{equation}
U_t^i=\mathbf{1}(D_t^i\geq L^i>D_{t-1}^i),
\end{equation}
with cumulative transmitted data \(D_t^i=\sum_{s=1}^t C_s^i\cdot\Delta t\), total load \(L^i\), and remaining data \(\phi_t^i=\max(L^i - D_t^i,0)\). The penalty term \(w>0\) discourages delays. The global reward \(R_t=\frac{1}{I}\sum_{i=1}^I r_t^i\) averages individual rewards, ensuring efficient system-wide resource use and connectivity.

\subsection{Multi-Agent Extension and State Estimation}

\begin{figure}
 \centering
 \includegraphics[width=8.8cm]{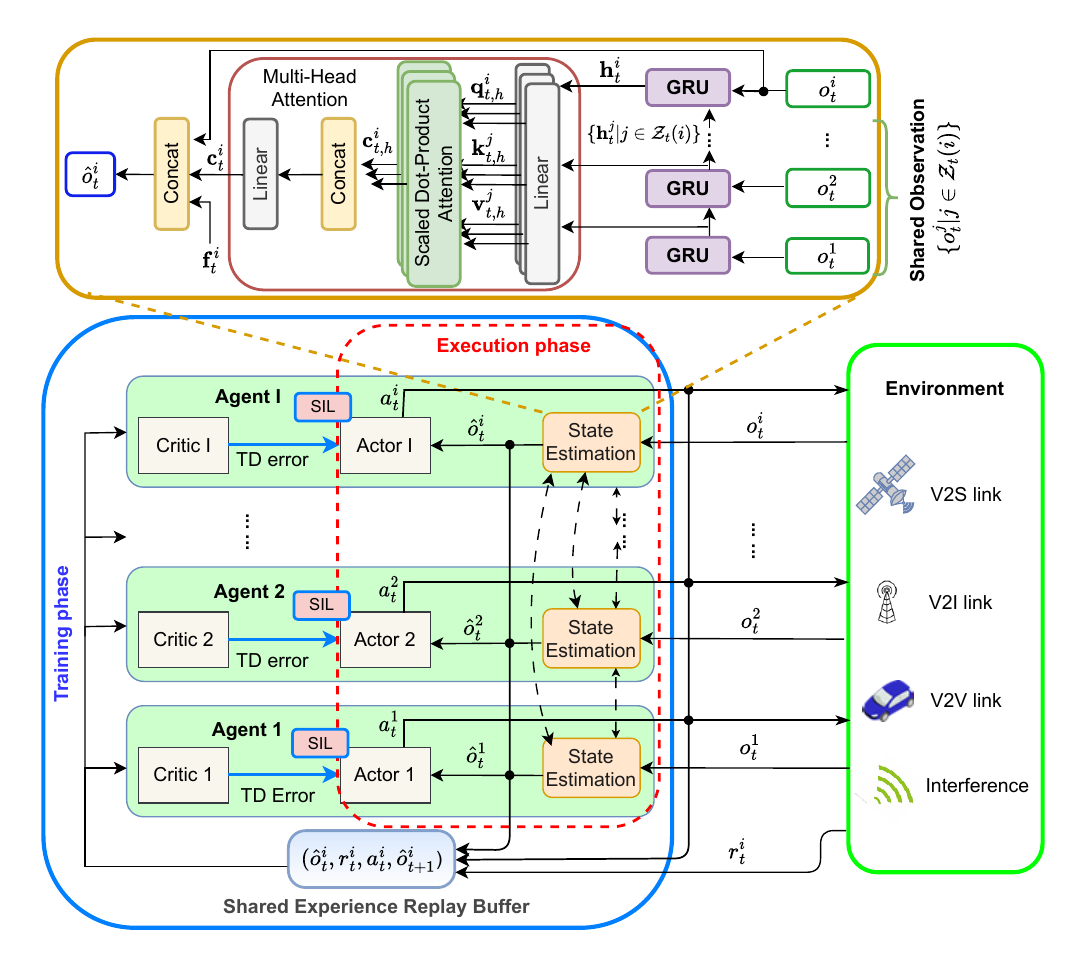}
 \caption{The proposed MAAC-SAM framework for vehicular networks.}
 \label{fig:MAAC-SAM}
\end{figure}

Fig.~\ref{fig:MAAC-SAM} illustrates the proposed MAAC-SAM framework, extending the traditional actor-critic method to decentralized satellite-aided vehicular networks with dynamic observation sharing.
 
Initially, variable-length observations from each agent \(i\) and its dynamic neighbor set \(\mathcal{Z}_t(i)\) are encoded by a GRU layer into fixed-size hidden states \(\mathbf{h}_t^i\) and \(\{\mathbf{h}_t^j|j \in \mathcal{Z}_t(i)\}\).

Subsequently, MHA integrates information from neighboring agents, refining the state estimate through its mechanism. For each attention head \( h \), query (\( \mathbf{q}_{t,h}^i \)), key (\( \mathbf{k}_{t,h}^j \)), and value (\( \mathbf{v}_{t,h}^j \)) vectors are linear transforms of GRU outputs:
\begin{equation}
\mathbf{q}_{t,h}^i = \mathbf{W}_h^Q \mathbf{h}_t^i,\quad
\mathbf{k}_{t,h}^j = \mathbf{W}_h^K \mathbf{h}_t^j,\quad
\mathbf{v}_{t,h}^j = \mathbf{W}_h^V \mathbf{h}_t^j,
\end{equation}
with learnable parameters \(\mathbf{W}_h^Q, \mathbf{W}_h^K, \mathbf{W}_h^V\). Attention weights use scaled dot-product attention:
\begin{equation}
\alpha_{t,h}^{i,j} = \frac{\exp\left(\frac{\mathbf{q}_{t,h}^{i\top}\mathbf{k}_{t,h}^{j}}{\sqrt{d_k}}\right)}{\sum_{m\in\mathcal{Z}_t(i)}\exp\left(\frac{\mathbf{q}_{t,h}^{i\top}\mathbf{k}_{t,h}^{m}}{\sqrt{d_k}}\right)},
\end{equation}
yielding context vectors per head:
\begin{equation}
\mathbf{c}_{t,h}^{i} = \sum_{j\in\mathcal{Z}_t(i)}\alpha_{t,h}^{i,j}\mathbf{v}_{t,h}^{j}.
\end{equation}
Concatenating outputs from \(H\) heads and applying an output projection \(\mathbf{W}^O\) yields \(\mathbf{c}_t^i\). This vector, combined with the agent's GRU hidden state \(\mathbf{h}_t^i\) and fingerprint features \(\mathbf{f}_t^i\) (training iteration, exploration rate~\cite{foerster2017stabilising}), produces an enhanced observation:
\begin{equation}
\hat{o}_t^i=\text{Concat}(\mathbf{h}_t^i,\mathbf{c}_t^i,\mathbf{f}_t^i).
\end{equation}
The enhanced observation \(\hat{o}_t^i\) informs the actor network, producing a policy \(\pi_{\theta^i}(a_t^i|\hat{o}_t^i)\), and the critic network, estimating value \(V_{\phi^i}(\hat{o}_t^i)\). Training is centralized for optimized learning, while execution remains decentralized to ensure scalability and adaptability.

Analyzing execution complexity for practical deployment, the state estimation network (GRU and MHA) complexity per step across agents is \(O(I(d_h^2 + d_h d_o + |\mathcal{Z}| d_h))\)\cite{vaswani2017attention}, where \(I\) is the agent count, \(d_h\) hidden dimension, \(d_o\) observation dimension, and \(|\mathcal{Z}|\) neighbor count per agent. Actor network complexity, a feedforward network, is \(O(\sum_{l=1}^{L} n_{l-1} n_l)\), with \(L\) layers, input dimension \(n_0=d_{\hat{o}}\), output dimension \(n_L\), and hidden layer sizes \(n_l\).

\subsection{Integrating Self-Imitation Learning}

The MAAC-SAM framework integrates SIL~\cite{oh2018self} to enhance MARL training efficiency in satellite-aided V2X communication. SIL explicitly encourages agents to revisit and reinforce successful historical experiences, accelerating policy improvement and enabling rapid adaptability in dynamic vehicular environments.

The SIL policy loss for agent \( i \) is defined as:
\begin{multline}
\mathcal{L}_{\text{SIL}}^{\pi^i}=-\frac{1}{M}\sum_{m=1}^{M}\log\pi_{\theta^i}(a^m|\hat{o}_t^m)\cdot A_\phi^+(a^m,\hat{o}_t^m)\\
-\frac{\beta}{M}\sum_{m=1}^{M}\mathcal{H}_t^{\pi^i}(\hat{o}_t^m),
\end{multline}
and the corresponding value loss is:
\begin{equation}
\mathcal{L}^{v^i}=\frac{1}{M}\sum_{m=1}^{M}\left[A_\phi(a^m,\hat{o}_t^m)\right]^2,
\end{equation}
Here, \(M\) represents the number of samples drawn from the agent's shared experience replay buffer, \(\hat{o}_t^m\) denotes the state estimate for sample \(m\) at time \(t\), and \(a^m\) is the corresponding action.
The advantage \(A_\phi(a^m, \hat{o}_t^m) = r^m + \delta V_\phi(\hat{o}_{t+1}^m) - V_\phi(\hat{o}_t^m)\) is the temporal difference (TD) error, where \(r^m\) is the reward, \(\delta\) is the discount factor.
The clipped advantage \(A_\phi^+(a^m, \hat{o}_t^m) = \max(0, A_\phi(a^m, \hat{o}_t^m))\) implements SIL and ensures only positive advantages influence the policy update. The entropy \(\mathcal{H}_t^{\pi^i}(\hat{o}_t^m) = - \sum_a \pi_{\theta^i}(a | \hat{o}_t^m) \log \pi_{\theta^i}(a | \hat{o}_t^m)\) with coefficient \(\beta\) is the regularization term to maintain exploration and prevent premature convergence.

\section{Numerical Results}\label{sec:results}

\subsection{Simulation Setup}

We comprehensively evaluated the MAAC-SAM framework using the realistic Simulation of Urban Mobility (SUMO) model~\cite{SUMO2018}, which simulates a wide range of traffic conditions. The range of vehicular density was selected according to previous studies \cite{rasheed2025deepbeam,xue2017roadside}. The simulations align explicitly with the latest 3GPP standards for terrestrial V2X\cite{3gpp38.886} and satellite-aided communication\cite{TR38.821}. Satellite positioning relies on accurate orbital data from Celestrak\cite{celestrak_starlink_data}. Key simulation parameters and hyperparameters optimized via Optuna\cite{akiba2019optuna} are detailed in Tables~\ref{table:system_parameters} and~\ref{tab:MAAC-A Parameters}. 

\begin{table}
\renewcommand{\arraystretch}{1.2}
\caption{Simulation Parameters\cite{3gpp38.886,TR38.821}}
\label{table:system_parameters}
\centering
 \scalebox{0.75}{%
\begin{tabular}{|p{4.5cm}|p{5.8cm}|}
\hline
\textbf{Parameter} & \textbf{Value} \\
\hline
Area (km x km) & 8.7 $\times$ 11.7 \\
\hline
Number of RSUs & 54 \\
\hline
Vehicular Density (Vehical/km$^2$) & 16.95, 25.42, 33.9, 42.37 \\
\hline
Carrier Frequency (GHz) & 3.5 \\
\hline
Antenna Height, Gain & Vehicle: 1.5 m, 3 dBi; BS: 25 m, 8 dBi \\
\hline
Noise Figure (dB) & Vehicle: 9; BS: 5; Satellite: 1.2 \\
\hline
Vehicle Speed (m/s) & 6.21 to 13.07 \\
\hline
Satellite Orbit, Altitude, Band & LEO, 550 km, Ka (30 GHz) \\
\hline
(Sub-channels (MHz)) & (V2V2I: $10 \times 1$; V2S: $20 \times 20$) \\
\hline
Packet Size (bytes) & V2V: 260; V2I, V2S: 500 \\
(Time Period (ms)) & ($\tau_V$ = 3; T = 100) \\
\hline
(Satellite Rx/Tx Gain) & (30.5 dBi / 43.2 dBi) \\
\hline
(Transmit Power (dBm)) & (V2V: [23, 10, 15, 17]; V2I: 23; V2S: 33.5) \\
\hline
Scintillation Loss (dB) & 2.2 \\
\hline
Fading & (Shadowing: Log-normal; Fast: Rayleigh) \\
\hline
\end{tabular}}
\end{table}

\begin{table}
\renewcommand{\arraystretch}{1.3}
\centering
\caption{MAAC-SAM Hyperparameters}
\scalebox{0.75}{%
\begin{tabular}{|p{2.4cm}|c|p{2.4cm}|c|}
\hline
\textbf{Parameter} & \textbf{Value} & \textbf{Parameter} & \textbf{Value} \\
\hline
Total Episode & 600 & Mini-batch Size & 64 \\
\hline
Time Step & 100 & Discounting Factor & 0.92 \\
\hline
Actor Learning Rate & 0.0001 & Critic Learning Rate & 0.009 \\
\hline
Entropy Factor & 0.058 & Hidden Units & (256, 224, 64,128 )\\
\hline
Number of Heads & 4 & Dropout Rate & 0.2 \\
\hline
\end{tabular}}
\label{tab:MAAC-A Parameters}
\end{table}

The proposed MAAC-SAM framework is implemented in PyTorch and compared with the following algorithms:
\begin{itemize}
\item D3QN-LS\cite{ji2023multi}: the original algorithm operates on terrestrial links only; to evaluate its behavior in an integrated setting, we developed a satellite-enabled variant named D3QN-LS-SA.
\item MAAC\cite{chen2023multi}: a multi-agent actor-critic framework without multi-head attention for state estimation and without advanced training techniques.
\item MA-DQN\cite{liang2019spectrum}: a multi-agent deep Q network for spectrum sharing between V2V and V2I communications in terrestrial networks.
\item GA\cite{brahmi2020genetic}: a GA-based method for resource allocation in satellite-aided V2X communication as an alternative to deep learning approaches. 
\end{itemize} 

\subsection{Performance Evaluation}

Table~\ref{tab:estimation_percentages} clearly illustrates the robustness of the state estimation process, which is critically essential under varying information-sharing scenarios. Even with substantially reduced observation sharing (40\%), the MAAC-SAM maintains a strong predictive capability (R\(^2\)= 0.8337, accuracy=0.8726), demonstrating that the combination of MHA and GRU encoding effectively manages partial observability, thus addressing privacy and bandwidth constraints realistically encountered in vehicular networks.

\begin{table}
\renewcommand{\arraystretch}{1.3}
\centering
\caption{Comparative State Estimation Across Sharing Levels.}
\label{tab:estimation_percentages}
\scalebox{0.8}{%
\begin{tabular}{cccccc}
\hline
\makecell{Sharing (\%)} & MSE & RMSE & MAE & R\(^2\) & Accuracy \\
\hline
100 & 0.1341 & 0.3662 & 0.1199 & 0.8620 & 0.8922 \\
80 & 0.1546 & 0.3932 & 0.1301 & 0.8467 & 0.8831 \\
60 & 0.1707 & 0.4132 & 0.1363 & 0.8344 & 0.8775 \\
40 & 0.1749 & 0.4182 & 0.1433 & 0.8337 & 0.8726 \\
\hline
\end{tabular}}
\end{table}

\begin{figure}
\centering
\includegraphics[width=0.9\linewidth]{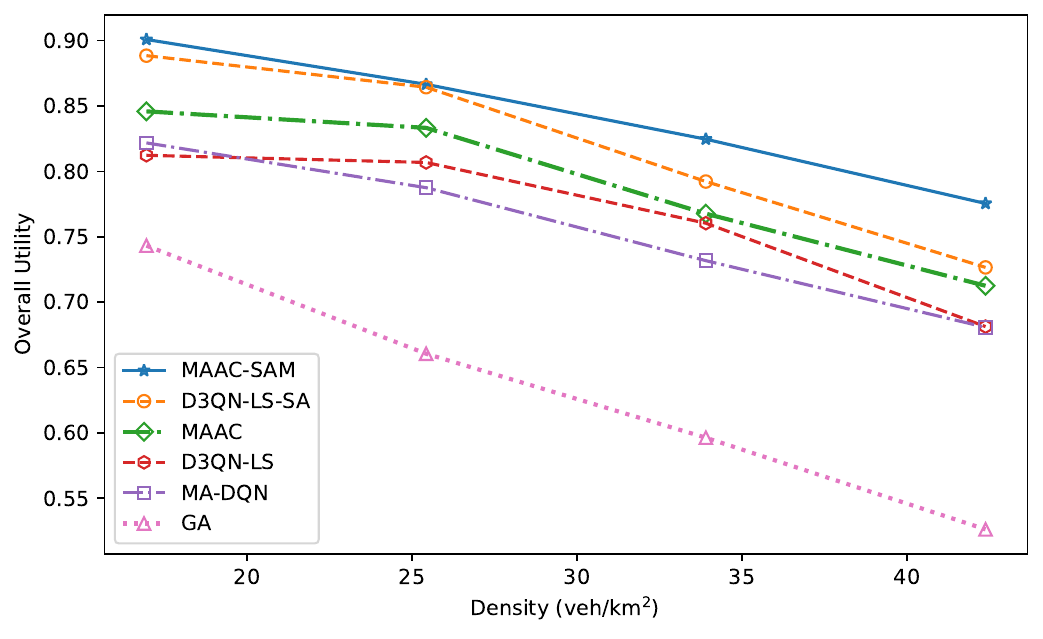}
 \caption{Utility comparison across vehicular densities.}
 \label{fig:utility_all_areas}
\end{figure}

Utility comparisons illustrated in Fig.~\ref{fig:utility_all_areas} underscore the consistent superiority of MAAC-SAM, especially evident in dense traffic conditions. The combination of precise MHA-based state estimation and rapid adaptability enabled by SIL allows MAAC-SAM to outperform terrestrial benchmarks (D3QN-LS, MA-DQN) by at least 14\% and maintain an 8\% lead over other satellite-aided algorithms. These results demonstrate the practical applicability of MAAC-SAM, showing substantial improvements in resource allocation and timely data transmission in realistic dynamic environments.

\begin{figure}
\centering
\includegraphics[width=0.85\linewidth]{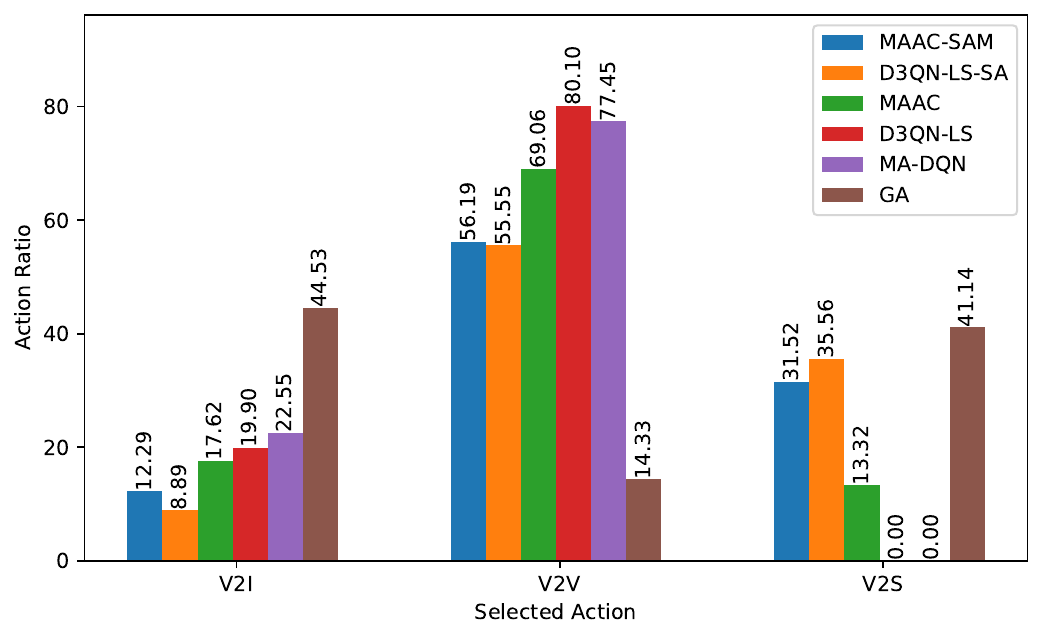}
 \caption{Channel Selection Distribution Across Communication Modes.}
 \label{fig:action_ratio}
\end{figure}

The strategic communication mode decisions of MAAC-SAM, as illustrated in Fig. \ref{fig:action_ratio}, emphasize its effective resource management beyond simple satellite access. By allocating 31.02\% of traffic to interference-free V2S links to avoid congestion while still utilizing V2V communication for 56.78\% of transmissions to balance terrestrial communications, MAAC-SAM ensures optimal resource utilization and dramatically enhances overall reliability. This flexibility demonstrates the practical benefits of integrating MHA and SIL for adaptive connectivity management.

Table~\ref{tab:utility_performance} presents an ablation study examining the performance of MAAC-SAM across different vehicle densities. The complete MAAC-SAM model demonstrates the highest utility. The elimination of fingerprinting (MAAC-SAM-NF) leads to a slight reduction in utility, suggesting that it plays a supportive role. In contrast, excluding SIL (MAAC-SAM-NoSIL) results in a notable decrease in performance, highlighting its critical importance in adapting to dynamic conditions. The baseline variant (MAAC), which excludes MHA, fingerprinting, and SIL, experiences the most significant performance drop, further emphasizing the value of the integrated MAAC-SAM framework for optimizing satellite-aided vehicular networks.

\begin{table}
\renewcommand{\arraystretch}{1.3}
\centering
\caption{Ablation Study on Key MAAC-SAM Components.}
\label{tab:utility_performance}
\scalebox{0.9}{%
\begin{tabular}{lcccc}
\hline
\multirow{2}{*}{\textbf{Method}} 
 & \multicolumn{4}{c}{\textbf{Utility under Various Vehicular Density}} \\ \cline{2-5}
 & 16.95 & 25.42 & 33.90 & 42.37 \\ \hline
MAAC-SAM & \textbf{0.9009} & \textbf{0.8666} & \textbf{0.8246} & \textbf{0.7754} \\ \hline
MAAC-SAM-NF & 0.8991 & 0.8620 & 0.8018 & 0.7676 \\ \hline
MAAC-SAM-NoSIL & 0.8667 & 0.8369 & 0.7786 & 0.7416 \\ \hline
MAAC & 0.8459 & 0.8332 & 0.7676 & 0.7124 \\ \hline
\end{tabular}%
}
\end{table}

\section{Conclusion}\label{sec:conclusion}

This paper introduces MAAC-SAM, a novel MARL framework designed for managing connectivity in satellite-aided vehicular networks. By integrating GRU-based encoders with a multi-head attention mechanism, MAAC-SAM effectively addresses the challenges posed by dynamic partial observability and fluctuating observation availability. Additionally, the implementation of SIL and fingerprinting significantly enhances the framework's learning efficiency, adaptability, and decision-making accuracy. Comprehensive simulation results confirm that MAAC-SAM significantly outperforms existing benchmarks, underscoring both its methodological innovation and practical relevance. Future work will focus on enhancing scalability and developing detailed channel modeling for practical, real-world deployment.

%
%

\bibliographystyle{IEEEtran}
\bibliography{ref}

\begin{thebibliography}{10}
\providecommand{\url}[1]{#1}
\csname url@samestyle\endcsname
\providecommand{\newblock}{\relax}
\providecommand{\bibinfo}[2]{#2}
\providecommand{\BIBentrySTDinterwordspacing}{\spaceskip=0pt\relax}
\providecommand{\BIBentryALTinterwordstretchfactor}{4}
\providecommand{\BIBentryALTinterwordspacing}{\spaceskip=\fontdimen2\font plus
\BIBentryALTinterwordstretchfactor\fontdimen3\font minus \fontdimen4\font\relax}
\providecommand{\BIBforeignlanguage}[2]{{%
\expandafter\ifx\csname l@#1\endcsname\relax
\typeout{** WARNING: IEEEtran.bst: No hyphenation pattern has been}%
\typeout{** loaded for the language `#1'. Using the pattern for}%
\typeout{** the default language instead.}%
\else
\language=\csname l@#1\endcsname
\fi
#2}}
\providecommand{\BIBdecl}{\relax}
\BIBdecl

\bibitem{He2021}
J.~He, K.~Yang, and H.-H. Chen, ``6g cellular networks and connected autonomous vehicles,'' \emph{IEEE Network}, vol.~35, no.~4, pp. 255--261, July/Aug 2021.

\bibitem{shang2024multi}
B.~Shang, X.~Li, Z.~Li, J.~Ma, X.~Chu, and P.~Fan, ``Multi-connectivity between terrestrial and non-terrestrial mimo systems,'' \emph{IEEE Open J. Commun. Soc.}, vol.~5, pp. 3245--3260, 2024.

\bibitem{gupta2025pora}
M.~S. Gupta, A.~Srivastava, and K.~Kumar, ``Pora: A proactive optimal resource allocation framework for spectrum management in cognitive radio networks,'' \emph{IEEE Transactions on Network and Service Management}, 2025.

\bibitem{deng2019joint}
B.~Deng, C.~Jiang, J.~Yan, N.~Ge, S.~Guo, and S.~Zhao, ``{Joint multigroup precoding and resource allocation in integrated terrestrial-satellite networks},'' \emph{IEEE Transactions on Vehicular Technology}, vol.~68, no.~8, pp. 8075--8090, 2019.

\bibitem{birabwa2022service}
D.~J. Birabwa, D.~Ramotsoela, and N.~Ventura, ``{Service-Aware User Association and Resource Allocation in Integrated Terrestrial and Non-Terrestrial Networks: A Genetic Algorithm Approach},'' \emph{IEEE Access}, vol.~10, pp. 104\,337--104\,357, 2022.

\bibitem{liang2019spectrum}
L.~Liang, H.~Ye, and G.~Y. Li, ``{Spectrum Sharing in Vehicular Networks Based on Multi-Agent Reinforcement Learning},'' \emph{IEEE Journal on Selected Areas in Communications}, vol.~37, no.~10, pp. 2282--2292, 2019.

\bibitem{ji2023multi}
Y.~Ji, Y.~Wang, H.~Zhao, G.~Gui, H.~Gacanin, H.~Sari, and F.~Adachi, ``{Multi-Agent Reinforcement Learning Resources Allocation Method Using Dueling Double Deep Q-Network in Vehicular Networks},'' \emph{IEEE Transactions on Vehicular Technology}, 2023.

\bibitem{chen2023multi}
P.-Y. Chen, Y.-H. Zheng, I.~Althamary, J.-L. Chern, and C.-W. Huang, ``{Multi-Agent Deep Reinforcement Learning for Spectrum Management in V2X with Social Roles},'' in \emph{IEEE Global Communications Conference (GLOBECOM)}, Kuala Lumpur, Malaysia, Dec 2023.

\bibitem{vaswani2017attention}
A.~Vaswani, N.~Shazeer, N.~Parmar, J.~Uszkoreit, L.~Jones, A.~N. Gomez, {\L}.~Kaiser, and I.~Polosukhin, ``{Attention is all you need},'' \emph{Advances in neural information processing systems}, vol.~30, 2017.

\bibitem{foerster2017stabilising}
J.~Foerster, N.~Nardelli, G.~Farquhar, T.~Afouras, P.~H. Torr, P.~Kohli, and S.~Whiteson, ``{Stabilising experience replay for deep multi-agent reinforcement learning},'' in \emph{International conference on machine learning}.\hskip 1em plus 0.5em minus 0.4em\relax PMLR, 2017, pp. 1146--1155.

\bibitem{oh2018self}
J.~Oh, Y.~Guo, S.~Singh, and H.~Lee, ``{Self-imitation learning},'' in \emph{International Conference on Machine Learning}.\hskip 1em plus 0.5em minus 0.4em\relax PMLR, 2018, pp. 3878--3887.

\bibitem{series2017propagation}
P.~Series, ``{Propagation data and prediction methods required for the design of Earth-space telecommunication systems},'' \emph{Recommendation ITU-R P.618-13}, 2017.

\bibitem{SUMO2018}
\BIBentryALTinterwordspacing
P.~A. Lopez, M.~Behrisch, L.~Bieker-Walz, J.~Erdmann, Y.-P. Fl{\"o}tter{\"o}d, R.~Hilbrich, L.~L{\"u}cken, J.~Rummel, P.~Wagner, and E.~Wie{\ss}ner, ``{Microscopic Traffic Simulation using SUMO},'' in \emph{The 21st IEEE International Conference on Intelligent Transportation Systems}.\hskip 1em plus 0.5em minus 0.4em\relax IEEE, 2018. [Online]. Available: \url{https://elib.dlr.de/124092/}
\BIBentrySTDinterwordspacing

\bibitem{rasheed2025deepbeam}
I.~Rasheed and H.~Mostafa, ``Deepbeam: A multi-agent deep reinforcement learning framework for predictive mmwave beam management in dynamic v2x networks,'' \emph{IEEE Transactions on Vehicular Technology}, 2025.

\bibitem{xue2017roadside}
L.~Xue, Y.~Yang, and D.~Dong, ``Roadside infrastructure planning scheme for the urban vehicular networks,'' \emph{Transportation Research Procedia}, vol.~25, pp. 1380--1396, 2017.

\bibitem{3gpp38.886}
3GPP, ``{V2X Services based on NR; User Equipment (UE) radio transmission and reception; (Release 16)},'' 3GPP, Tech. Rep., March 2021.

\bibitem{TR38.821}
------, ``{3GPP TR 38.821: Solutions for NR to support non-terrestrial networks (NTN)},'' 3GPP, Tech. Rep., April 2023.

\bibitem{celestrak_starlink_data}
{North American Aerospace Defense Command (NORAD)}, ``{Starlink Satellite Elements},'' Online, Jul. 2025, available: \url{http://celestrak.com/NORAD/elements/}. [Accessed: July 3, 2025].

\bibitem{akiba2019optuna}
T.~Akiba, S.~Sano, T.~Yanase, T.~Ohta, and M.~Koyama, ``Optuna: A next-generation hyperparameter optimization framework,'' in \emph{Proceedings of the 25th ACM SIGKDD international conference on knowledge discovery \& data mining}, 2019, pp. 2623--2631.

\bibitem{brahmi2020genetic}
I.~Brahmi, H.~Koubaa, and F.~Zarai, ``{Genetic algorithm based resource allocation for V2X communications},'' in \emph{2020 IEEE Eighth International Conference on Communications and Networking (ComNet)}.\hskip 1em plus 0.5em minus 0.4em\relax IEEE, 2020, pp. 1--5.

\end{thebibliography}

\end{document}